\documentclass[a4paper,11pt]{article}
\usepackage{pos}

\usepackage{slashed}
\usepackage{hyperref}
\usepackage{amsthm}
\usepackage{mathtools}
\usepackage{graphicx}
\usepackage{amsmath}
\usepackage{amsfonts}

\newcommand{\frg}{\mathfrak{g}}
\newcommand{\id}{\mathrm{id}}
\newcommand{\CF}{\mathcal{F}}
\newcommand{\CCL}{\mathscr{L}}
\newcommand{\CR}{\mathcal{R}}
\newcommand{\CA}{\mathcal{A}}
\newcommand{\ii}{{\mathrm{i}}}
\newcommand{\frv}{\mathfrak{v}}
\newcommand{\sfR}{\mathrm{R}}
\newcommand{\Tr}{\mathrm{Tr}}
\newcommand{\nn}{\nonumber}
\def\d{{\rm d}}
\def\ds{\stackrel{\star}{,}}
\def\kbar{{\mathchar'26\mkern-9muk}}
\def\RR{{\mathcal R}}

\title{$L_\infty$-algebra of braided electrodynamics}

\author*[a]{M. Dimitrijevi\' c \'Ciri\'c}
\author[a]{N. Konjik}
\author[a]{V. Radovanovi\' c}
\author[b]{R. J. Szabo}
\author[a]{M. Toman}

\affiliation[a]{University of Belgrade, Faculty of Physics\\
Studentski trg 12, Belgrade, Serbia}

\affiliation[b]{Department of Mathematics, Heriot-Watt University, Edinburgh, United
Kingdom\\
Maxwell Institute for
Mathematical Sciences, Edinburgh, United Kingdom\\
Higgs Centre
for Theoretical Physics, Edinburgh, United Kingdom}

\emailAdd{dmarija@ipb.ac.rs}
\emailAdd{konjik@ipb.ac.rs}
\emailAdd{rvoja@ipb.ac.rs}
\emailAdd{R.J.Szabo@hw.ac.uk}
\emailAdd{misa.toman97@gmail.com}

\abstract{Using the recently developed formalism of braided noncommutative field theory, we construct an explicit example of braided electrodynamics, that is, a noncommutative $U(1)$ gauge theory coupled to a Dirac fermion. We construct the braided $L_\infty$-algebra of this field theory and apply the formalism to obtain the braided equations of motion, action functional and conserved matter current. The braided deformations leads to a modification of the charge conservation. Finally, the Feynman integral appearing in the one-loop contribution to the vacuum polarization diagram is calculated. There are no non-planar diagrams, but the UV/IR mixing appears nevertheless. We comment on this unexpected result.}

\notes{\note{Preprint:  {\sf EMPG--22--07}}}

\FullConference{%
  Corfu Summer Institute 2021 "School and Workshops on Elementary Particle Physics and Gravity"\\
  29 August - 9 October 2021\\
  Corfu, Greece
}

\begin{document}
\maketitle

\section{Introduction}

\noindent $L_\infty$-algebras are generalizations of differential graded Lie algebras with infinitely-many graded antisymmetric
brackets, related to each other by higher homotopy versions of the
Jacobi identity. In~\cite{HohmZwiebach} it was suggested that the complete data of
any classical field theory with generalized gauge symmetries fit into cyclic $L_\infty$-algebras
with finitely-many non-vanishing brackets, encoding both gauge
transformations and dynamics. It was then showed that the existence of such an
$L_{\infty}$-algebra formulation is a consequence of the duality with the BV--BRST
formalism for perturbative field theories \cite{BVChristian}.\smallskip

\noindent $L_\infty$-algebras naturally appear in noncommutative gauge theory \cite{Munich18, Kup1, Kup2, Kupriyanov:2021cws} and noncommutative gravity \cite{BraidedLinf} (see also the contribution \cite{Szabo:2022edp} to these proceedings for a brief review). In \cite{Munich18} it was shown that the semi-classical limit of a noncommutative and/or nonassociative gauge theory can be encoded in an infinite dimensional $L_\infty$-algebra which is constructed order by order in the deformation parameter. Furthermore, it was shown in \cite{Kup2} that the Seiberg-Witten map relating noncommutative gauge theory with the corresponding commutative gauge theory is a $L_\infty$ quasi-isomorphism. Using the Drinfel'd twist deformation method in our recent work \cite{BraidedLinf, RichardGregory} we constructed a deformation of a $L_\infty$-algebra, the braided $L_\infty$-algebra. The corresponding field theory is then the braided gauge theory. It is different compared to the usual noncommutative gauge theory, the $\star$-gauge theory. The braided gauge transformations close in the Lie algebra of the undeformed gauge symmetry and they have a braided Leibniz rule.\smallskip

\noindent
In this paper we illustrate our construction of braided $L_\infty$-algebra and braided gauge theories on the example of braided electrodynamics: braided $U(1)$ gauge theory coupled to a charged Dirac spinor. We start by reviewing some basic facts about $L_\infty$-algebras and their relation with classical field theories. Then we introduce the braided $L_\infty$-algebra and the corresponding braided gauge theory. To illustrate our construction, we discuss in details the braided electrodynamics and its properties. In particular, the theory remains abelian and there are no three and four photon vertices. The quantization of field theories with braided symmetries is currently under development. To gain some preliminary insight, here we calculate the standard Feynman integrals which appear in the one-loop contribution to the vacuum polarization. We find UV/IR mixing, but no non-planar diagrams. This unexpected result should be understood once the full quantum field theory of braided (gauge) field theories is constructed. Some preliminary results on these problems can be found in \cite{SzaboAlex, Oeckl}.

\section{$L_\infty$-algebras and classical field theory}

\noindent In this section we briefly review the connection between classical field theories and $L_\infty$-algebras established in \cite{HohmZwiebach, BVChristian}.\smallskip

\noindent An $L_\infty$-algebra is a $\mathbb{Z}$-graded vector space {\small$V=\bigoplus_{k\in 
\mathbb{Z}}\, V_{k}$} with graded antisymmetric multilinear maps called $n$-brackets
\begin{align}
\ell_{n}:  \text{\Large$\otimes$}^{n}V \longrightarrow & V \ , \quad  a_1\otimes \cdots\otimes a_n \longmapsto 
\ell_{n} 
(a_1,\dots,a_n) \nn\\
\ell_{n} (\dots, a,a',\dots) &=\>  -(-1)^{|a|\,|a'|}\, \ell_{n} (\dots, a',a,\dots) \ ,\nn
\end{align}
\noindent where $|a|$ is the degree of a homogeneous element $a\in V$. The $n$-brackets must also fulfil homotopy relations. The first three relations are given by
\begin{align}\label{HomRelations}
n=&1:  \quad \ell_{1}\big(\ell_{1}(a)\big) = 0,\nn\\
n=&2:  \quad \ell_{1}\big(\ell_{2}(a_1,a_2)\big) = \ell_{2}\big(\ell_{1}(a_1),a_2\big) + (-1)^{|a_1|}\, 
\ell_{2}\big(a_1, 
\ell_{1}(a_2)\big)\ ,\\
n=&3: \quad  \ell_{1}\big(\ell_{3}(a_1,a_2,a_3)\big) =  - \ell_{3}\big(\ell_{1}(a_1),a_2,a_3\big) - (-1)^{|a_1|}\, 
\ell_{3}\big(a_1, \ell_{1}(a_2), a_3\big) \nn\\
& - (-1)^{|a_1|+|a_2|}\, \ell_{3}\big(a_1,a_2, \ell_{1}(a_3)\big) \nn\\
&-\ell_{2}\big(\ell_{2}(a_1,a_2),a_3\big) -(-1)^{(|a_1|+|a_2|)\,|a_3|}\, \ell_{2}\big(\ell_{2}(a_3,a_1),a_2\big) \nn\\
& -(-1)^{(|a_2|+|a_3|)\,|a_1|}\, \ell_{2}\big(\ell_{2}(a_2,a_3),a_1\big) \nn
\end{align}

\noindent Cyclic $L_\infty$-algebras contain an additional structure called cyclic pairing that is a graded, symmetric, and non-degenerate bilinear map $\langle-,-\rangle:V\otimes V\to\mathbb{R}$ satisfying
\begin{align}
\langle a_0,\ell_n(a_1,a_2,\dots,a_n)\rangle = (-1)^{n+(|a_0|+|a_n|)\,n+|a_n| \,\sum_{i=0}^{n-1}\,|a_i|} \ \langle 
a_n,\ell_n(a_0,a_1,\dots,a_{n-1})\rangle , \>\>  n\geq 1.\label{Pairing}
\end{align}

\noindent In order to relate this formalism to a (gauge) field theory, we first define the graded vector space $V=V_{0}\oplus V_{1}\oplus V_{2}\oplus V_{3}$. This space contains gauge parameters $\rho \in V_{0}$, gauge fields $A\in V_{1}$, equations of motion $F_A\in V_{2}$, and II Noether identities $\d_A F_A\in V_{3}$. The gauge transformations $\delta_\rho A$, equations of motion $F_A = 0$, gauge invariant action functional $S$ and the second Noether identity $\d_A F_A =0$ are then formulated as follows:
\begin{align}
\delta_{\rho}A &= \ell_1(\rho) + \ell_2(\rho,A) -\frac{1}{2}\ell_3(\rho,A, A) + \dots,\label{eq:gaugetransfA}\\
F_A  &= \ell_1(A) - \frac{1}{2}\ell_2(A,A) -\frac{1}{3!}\ell_3(A,A,A) +\dots , \label{eq:eom}\\
S(A)  &= \frac{1}{2} \langle A, \ell_{1}(A)\rangle - \frac{1}{3!}\langle A, \ell_{2}(A,A)\rangle + \dots, \label{eq:action}\\
\d_A F_A &= \ell_1(F_A) + \ell_2(F_A,A) +\dots\label{eq:Noether}
\end{align}

\noindent In field theories the cyclic pairing of degree $-3$ is needed. Therefore, the only non-vanishing pairings are 
\begin{equation}
\langle-,-\rangle:V_k\otimes V_{3-k}\longrightarrow \mathbb{R} \qquad
\mbox{for} \quad k\leq 3 \ . \nn
\end{equation}
The variation principle is then written as
\begin{equation}
\delta S(A) = \langle \delta A\, ,\, F_A\rangle \ .\nn
\end{equation}
We will now illustrate this $L_\infty$-algebra encoding using two important examples.

\vspace{5mm}
\noindent {\bf $L_\infty$-algebra of 3D nonabelian Chern-Simons theory}

\vspace{3mm}
\noindent Consider a Chern-Simons theory in three dimensions for a gauge group $G$ with the corresponding (hermitian) Lie algebra generators $T^a$, $a=1\dots n$, whose Lie algebra has an invariant quadratic form $\textrm{Tr}$ and commutation relations $\left[A^a, A^b\right] = i f^{abc} T^c$. Let $A$ be a Lie algebra valued one-form, $A = A_\mu\d x^\mu$, transforming under gauge transformations as $\delta_\rho A = \d\rho + i\left[\rho, A\right]$. The curvature tensor is $F = \d A - i A\wedge A = \d A - \frac{i}{2}\left[A, A\right].$ The action of this theory can be written as
\begin{align*}
S & = \frac{1}{2}\int\Tr\left(A\wedge\d A - \frac{i}{3} A\wedge\left[A, A\right]\right),
\end{align*}
which yields the equation of motion $F = 0$.

\noindent This theory can be encoded into the $L_\infty$-algebra formalism by introducing $V = V_0\oplus V_1\oplus V_2\oplus V_3$. The corresponding nonvanishing $\ell$-brackets are given by
\begin{align}
& \ell_1(\rho) = \d\rho, \quad \ell_2\left(\rho_1, \rho_2\right) = i\left[ \rho_1, \rho_2\right], \quad \ell_2\left(\rho, A\right) = i\left[ \rho, A\right], \nn\\
& \ell_1\left(A\right) = \d A,\quad \ell_2\left(A_1, A_2\right) = i[A_1, A_2],\nn\\
& \ell_1\left(F_A\right)  = \d F_A, \quad \ell_2\left(A, F_A\right) = i[A, F_A].\nn
\end{align}
In addition, a cyclic pairing of degree $-3$ is defined as
\begin{equation}
\langle \alpha, \beta \rangle = \int \textrm{Tr}\, \alpha\wedge \beta ,\nn
\end{equation}
where $\alpha$ and $\beta$ are Lie algebra valued differential forms. Since the cyclic pairing is a map of degree $-3$, it is non-vanishing only on the homogeneous subspaces $V_0\otimes V_3$ and $V_1\otimes V_2$.

\noindent The theory is reproduced as
\begin{align*}
\delta_\rho A & = \ell_1\left(\rho\right) + \ell_2\left(\rho, A\right)  = \d\rho + i\left[\rho, A\right],\\
F_A  &= \ell_1(A) - \frac{1}{2}\ell_2(A,A) = \d A -i A\wedge A,\\
\d_A F_A & = \ell_1\left(F_A\right) - \ell_2\left(F_A, A\right) = \d F_A -i \left[A, F_A\right],
\end{align*}

\noindent The action can be written with the help of cyclic pairing as
\begin{align*}
S(A) &= \frac{1}{2} \langle A,\ell_1(A) \rangle - \frac{1}{3!} \langle A, \ell_2(A, A)\rangle = \frac{1}{2}\int\textrm{Tr}\left(A\wedge\d A - \frac{i}{3}A\wedge\left[A, A\right]\right).
\end{align*}

\vspace{5mm}
\noindent {\bf $L_\infty$ algebra of 4D $\phi^4$ theory}
\vspace{3mm}

\noindent This theory is not a gauge theory and therefore $V= V_{1}\oplus V_{2}$. The corresponding nonvanishing brackets are given by
\begin{align*}
\ell_1\left(\phi\right) & = \left(\Box - m^2\right)\phi,\quad \ell_3\left(\phi_1, \phi_2, \phi_3\right) = \lambda \phi_1 \phi_2 \phi_3.
\end{align*}

\noindent The cyclic pairing can be taken to be
\begin{align*}
\langle\phi_1, \phi_2\rangle & = \int\d^4 x\phi_1(x)\phi_2(x)
\end{align*}
where $\phi_1\in V_1$ and $\phi_2\in V_2$. This leads to the action functional
\begin{align*}
S(\phi) & = \frac{1}{2} \langle \phi, \ell_1\left(\phi\right)\rangle - \frac{1}{4!} \langle\phi, \ell_3\left(\phi, \phi, \phi\right)\rangle \\
& = \int\d^4 x\left(\frac{1}{2}\phi\left(\Box - m^2\right)\phi - \frac{\lambda}{4!}\phi^4\right).
\end{align*}

\noindent The equation of motion is
\begin{align*}
F_\phi & = \ell_1(\phi) - \frac{1}{2}\ell_2(\phi, \phi) -\frac{1}{3!}\ell_3(\phi, \phi, \phi)  = \left(\Box - m^2\right)\phi - \frac{\lambda}{3!}\phi^3 = 0
\end{align*}

\section{Braided gauge theory and its braided $L_\infty$-algebra}

\noindent In this section we first briefly review the recently proposed description of the noncommutative gauge theory: the braided gauge theory. Then we relate this theory with the notion of braided $L_\infty$-algebra. More details can be found in \cite{BraidedLinf, RichardGregory}.

\vspace{5mm}  
\noindent{\bf Braided gauge theory}
\vspace{3mm}

\noindent Noncommutative deformation of gauge theories can be defined in different ways. One of the most studied examples is that of $\star$-gauge theories \cite{Szabo:2001kg, Star-gauge}. Let us again use the gauge group $G$ with the corresponding hermitian Lie algebra generators $T^a$, $a=1\dots n$, the invariant quadratic form $\textrm{Tr}_\frg$ and commutation relations $\left[A^a, A^b\right] = i f^{abc} T^c$. Let the gauge field $A$ be a Lie algebra valued one-form, $A = A^a_\mu T^a\d x^\mu$\footnote{Note that we work with the matrix Lie algebra $\frg$.}. An infinitesimal gauge transformation is defined as
\begin{equation}
\tilde{\delta}_{\rho}^{\star}A = \d\rho + i[\rho \ds A]  = \d\rho + i \big( \rho\star A - A\star \rho  \big)\  \nn 
\end{equation}
with the undeformed Leibniz rule
\begin{equation}
\triangle(\tilde{\delta}^\star_\rho) = \tilde{\delta}^\star_\rho\otimes\id + \id\otimes\tilde{\delta}^\star_\rho  \ . \nn
\end{equation}
However, it is easily checked that
\begin{equation}
[\rho^a T^a \ds A^b T^b] = \frac{1}{2}\{\rho^a \ds A^b \} [T^a, T^b] + \frac{1}{2}[\rho^a \ds A^b ] \{T^a, T^b\} .\nn
\end{equation}
Therefore, in general these transformations do not close in the corresponding Lie algebra. To 
circumvent this problem, one either works with the $u(n)$ algebra in its fundamental representation, or enlarges the algebra to the universal enveloping algebra. Working with the universal enveloping algebra results in infinitely many new degrees of freedom, and one might then use the Seiberg-Witten map to express all of the new degrees of freedom in terms of the corresponding classical (commutative) degrees of freedom \cite{JSSW}.

\noindent Let us now define a noncommutative gauge theory using the notion of a braided Lie algebra \cite{SLN}. A short review of the twist formalism, $R$ matrix and the $\star$-products is presented in Appendix~\ref{app:Drinfeld}. For simplicity, we consider the example of braided Chern-Simons non-abelian gauge theory in $3D$. The gauge field $A = A^a_\mu T^a\d x^\mu$ transforms as ($\rho =\rho^aT^a$)
\begin{equation}
\delta_{\rho}^{\star}A = \d\rho + i[\rho,A]_\star  = \d\rho + i \big( \rho\star A - \sfR_k(A)\star \sfR^k (\rho)  \big)\ . \label{BrGTransformation}
\end{equation}
It is easily verified that the braided commutator closes in the Lie algebra
\begin{equation}
[\rho_1 , \rho_2]_\star = if^{abc}\rho_1^b\star\rho_2^c T^a.\nn 
\end{equation}

\noindent Transformations (\ref{BrGTransformation}) have the braided Leibniz rule 
\begin{equation}
\triangle_\CF(\delta^\star_\rho) = \delta^\star_\rho\otimes\id +
\sfR_k\otimes\delta^\star_{\sfR^k(\rho)  \ , } \label{BrLRule}
\end{equation}
and close the braided algebra
\begin{align}
\big[\delta_{\rho_1}^{\star}, \delta_{\rho_2}^{\star}\big]_\circ^\star \>=&  \delta_{\rho_1}^{\star}\circ \delta_{\rho_2}^{\star} - \delta_{\sfR_k(\rho_2)}^{\star}\circ \delta_{\sfR^k(\rho_1)}^{\star} = \delta_{-i[\rho_1, \rho_2]_\star}^{\star} \ .\nn
\end{align}
Note that in this setting we can define left and right gauge transformations
\begin{equation}
\delta_{\rho}^{\star{\mbox{\tiny L}}}A = \d\rho + i[\rho,A]_\star \> {\mbox{ and }} \> \delta_{\rho}^{\star{\mbox{\tiny R}}}A = \d\rho - i[A, \rho]_\star .\nn
\end{equation}
They are different. We will work only with left gauge transformations, analogous conclusions also hold for the right gauge transformations, see \cite{BraidedLinf} for more details.

\noindent The braided curvature of the gauge field $A$ is given by
\begin{equation}
F_A^\star =  \d A - i\frac12\, [A,A]_\star = \big( \d A^a + \frac{1}{2}f^{abc} A^b\wedge_\star A^c \big)T^a \ , \nn
\end{equation}
and transforms covariantly
\begin{equation}
\delta_\rho^{\star}F_A^\star  = i[\rho,F_A^\star]_\star \ .\nn 
\end{equation}
The gauge invariant action is defined as
\begin{equation}
S_\star(A) = \frac12\, \int_M\, \Tr_\frg\Big(A\wedge_\star\d A 
- \frac{i}{3}\, A\wedge_\star [A, A]_\star\Big) \ . \nn
\end{equation}
Unlike in the commutative case, the braided second Noether identity is not linear in field equations
\begin{align}
\d_A^\star F_A^\star = & \>\d F_A^\star - \frac{i}{2}\,[A,F_A^\star]_\star - \frac{i}{2}\,[F_A^\star,A]_\star + \tfrac14\,[\sfR_k(A),[\sfR^k(A),A]_\star]_\star = 0  .\nonumber
\end{align}
There is an additional term on the right-hand side that depends only on the gauge field.

\vspace{5mm}
\noindent{\bf Braided $L_\infty$-algebra}
\vspace{3mm}

\noindent Starting with a suitable classical $L_\infty$-algebra, $\CCL$, a braided $L_\infty$-algebra, $\CCL^\star$, can be constructed using Drinfel'd twist deformation techniques as in~\cite{BraidedLinf}. More details on the Drinfel'd twist deformation formalism can be found in \cite{SLN} and \cite{MajidBook}.

\noindent Following the prescription outlined in Appendix~\ref{app:Drinfeld}, we set the first bracket $\ell_1^\star:=\ell_1$ and
\begin{align}\label{eq:ellnstardef}
\ell_n^\star(a_1,\dots,a_n) :=
\ell_n(a_1\otimes_\star\cdots\otimes_\star a_n)
\end{align}
for $n\geq2$, where $a\otimes_\star a':=\CF^{-1}(a\otimes
a')=\bar{\mathrm{f}}^k(a)\otimes\bar{\mathrm{f}}_k(a')$ for $a,a'\in V$. These define multilinear maps $\ell_n^\star: V^{\otimes n}\to V$ which are braided graded antisymmetric:
\begin{align*}
\ell_n^\star (\dots, a,a',\dots) = -(-1)^{|a|\,|a'|}\, \ell_n^\star \big(\dots,
  \sfR_k(a'),\sfR^k(a),\dots\big) \ .
\end{align*}

\noindent The first and second homotopy relations are unchanged with respect to the corresponding classical homotopy relations, that is, the braided $L_\infty$-algebra $\CCL^\star$ still has underlying cochain complex $(V,\ell_1)$ and $\ell_2^\star$ is again a cochain map. The third homotopy relation is given by
\begin{align}
\begin{split}
& \ell^\star_2\big(\ell^\star_2(a_1,a_2),a_3\big) - (-1)^{|a_2|\,|a_3|}\,
  \ell^\star_2\big(\ell^\star_2(a_1,\sfR_k(a_3)),\sfR^k(a_2)\big) \\ 
&\hspace{6.5cm} +  (-1)^{(|a_2|+|a_3|)\,|a_1|}\,
  \ell^\star_2\big(\ell^\star_2(\sfR_k(a_2),\sfR_l(a_3)),\sfR^l\,\sfR^k(a_1)\big) \\[4pt]
& \hspace{1.5cm} = -\ell^\star_3\big(\ell_1(a_1),a_2,a_3\big) - (-1)^{|a_1|}\,
  \ell^\star_3\big(a_1, \ell_1(a_2), a_3\big) - (-1)^{|a_1|+|a_2|}\,
  \ell^\star_3\big(a_1,a_2, \ell_1(a_3)\big) \\
& \hspace{6.5cm} -\ell_1\big(\ell^\star_3(a_1,a_2,a_3)\big) \ .
\end{split}
\label{I3braided} 
\end{align}
We see that the non-trivial braiding now appears in this relation, which indicates that the braided graded Jacobi identity for $\ell_2^\star$ is violated by the cochain homotopy $\ell_3^\star$. Nontrivial braiding also appears in the higher homotopy relations \cite{BraidedLinf}.

\noindent The graded symmetry of the cyclic pairing $\langle-,-\rangle$
implies that the twisted pairing $\langle-,-\rangle_\star$ is naturally
braided graded symmetric 
\begin{equation}
\langle a_1,a_2\rangle_\star  = \langle \bar{\mathrm{f}}^k(a_1), \bar{\mathrm{f}}_k(a_2)\rangle = (-1)^{|a_1|\,|a_2|} \, \langle \sfR_k(a_2),\sfR^k(a_1)\rangle_\star \  .  \label{CyclPairingStar}
\end{equation}
However, for applications to field theory, we have to restrict to compatible Drinfel'd twists~\cite{BraidedLinf} that result in a strictly graded symmetric pairing
\begin{align*}
\langle
  a_1,a_2\rangle_\star  = (-1)^{|a_1|\,|a_2|} \, \langle
  a_2,a_1\rangle_\star 
\end{align*}
for all homogeneous $a_1,a_2\in V$. In this case, $\CCL^\star$ becomes a strictly cyclic braided $L_\infty$-algebra.

\noindent Let $\CCL^\star = (V,\{\ell^\star_n\})$ be a $4$-term braided $L_\infty$-algebra, obtained by twist deformation of an $L_\infty$-algebra $\CCL=(V,\{\ell_n\})$, which organizes the symmetries and dynamics of a classical field theory. For a gauge parameter $\rho\in V_0$, we define the braided gauge variation of a field $A\in V_1$ by
\begin{align}\label{eq:LgtL} 
\delta_\rho^\star A = \ell_1(\rho) + \ell_2^\star(\rho,A) - \tfrac12\,\ell_3^\star(\rho,A,A) + \cdots
 \ .
\end{align}
Braided covariant dynamics are described by the equations of motion $F_A^\star=0$
\begin{align}\label{eq:braidedeom}
F_A^\star =  \ell_1(A) - \tfrac12\,\ell_2^\star(A,A) - \tfrac1{6}\,\ell_3^\star(A,A,A) + \cdots 
\end{align}
that transform covariantly as
\begin{align}\label{eq:lefteomcov}
\delta_\rho^\star F_A^\star = \ell_2^\star(\rho,F_A^\star) + \tfrac12\,\big(\ell_3^\star(\rho,F_A^\star,A) - \ell_3^\star(\rho,A,F_A^\star)\big) + \cdots   \ ,
\end{align}
for all gauge parameters $\rho \in V_0$. 

\noindent For the field theories considered in this paper, the braided gauge transformations obey the off-shell closure relation in terms of the braided commutator:
\begin{align}\label{eq:braidedclosure}
\big[\delta_{\rho_1}^\star,\delta_{\rho_2}^\star\big]_\circ^\star :=  \delta_{\rho_1}^\star\circ\delta_{\rho_2}^\star - \delta_{\sfR_k(\rho_2)}^\star\circ\delta_{\sfR^k(\rho_1)}^\star
= \delta_{-\ell^\star_2(\rho_1,\rho_2)}^\star \ .
\end{align}

\noindent Corresponding to the braided gauge symmetry, a suitable combination of the braided homotopy relations leads to an identity
\begin{align}\label{eq:braidedNoether}
\d^\star_A F_A^\star :&\! = \ell_1(F_A^\star) + \tfrac12\,\big(\ell^\star_2(F_A^\star,A) - \ell^\star_2(A,F_A^\star)\big) +  \tfrac1{6} \, \ell_1\big(\ell^\star_{3}(A,A,A)\big)+ \cdots\nonumber  \\
& \hspace{1cm} \quad \, + \tfrac18\,\big(\ell_2^\star(\ell_2^\star(A,A),A) - \ell_2^\star(A,\ell_2^\star(A,A))\big) \nonumber\\
& \hspace{3cm} \quad \, + \tfrac1{12}\,\big( \ell_2^\star(\ell_3^\star(A,A,A),A) - \ell_2^\star(A,\ell_3^\star(A,A,A))\big) + \cdots \ = \ 0 \ . 
\end{align}
We have already seen in the example of the braided Chern-Simons theory that, unlike the classical Noether identity \eqref{eq:Noether}, the braided Noether identity (\ref{eq:braidedNoether}) is no longer linear in the equations of motion $F^\star_A$ and contains inhomogeneous terms involving brackets of the fields $A$ themselves. This is related to the violations of the Bianchi identities in braided gauge theories~\cite{BraidedLinf}. In the classical limit, where $\CR=1\otimes1$, the braided Noether identity \eqref{eq:braidedNoether} reduces to the classical formula \eqref{eq:Noether}.

\noindent For a Lagrangian field theory, using the (strictly) cyclic inner product one can define an analogue of the action functional for the braided field theory as
\begin{align}\label{eq:braidedaction}
S_\star(A) := \tfrac12\,\langle A,\ell_1(A)\rangle_\star - \tfrac16\,\langle A,\ell^\star_2(A,A)\rangle_\star - \tfrac1{24}\,\langle A,\ell^\star_3(A,A,A)\rangle_\star +\cdots \ ,
\end{align}
whose variational principle yields the braided equations of motion $F_A^\star=0$.
This action functional is invariant under braided gauge transformations:
\begin{align}\label{eq:braidedgaugeinv}
\delta_\rho^\star S_\star(A)=0 \ ,
\end{align}
for all $\rho\in V_0$ and $A\in V_1$. Note that the free fields of braided field theory are unchanged from the classical field theory. Only the interaction vertices, corresponding to the higher brackets $\ell_n^\star$ for $n\geq2$, are modified by the braided noncommutative deformation.

\section{Braided electrodynamics}

\noindent In this section we first rewrite the classical electrodynamics as an $L_\infty$-algebra. More examples of $L_\infty$-algebra description of classical gauge theories coupled to matter are discussed in \cite{LInfMatter}. Following the steps from Section 3 we formulate a noncommutative generalization and obtain a braided $L_\infty$-algebra of noncommutative electrodynamics. For simplicity we will work with the Moyal-Weyl deformation and do all calculations in the coordinate basis. For a more general result we refer to \cite{UsNew}.

\vspace{5mm}
\noindent {\bf $L_\infty$-algebra of classical electrodynamics}
\vspace{3mm}

\noindent The classical electrodynamics on the 4D Minkowski space-time is a $U(1)$ gauge theory with the massive spinor field $\psi$, and the $U(1)$ gauge field $A_\mu$. The infinitesimal $U(1)$ gauge transformations are
\begin{equation}
\delta_\rho \psi = i\rho\psi, \quad \delta_\rho \bar{\psi} = -i\bar{\psi}\rho  ,\quad \delta_\rho A_\mu = \frac{1}{e}\partial_\mu\rho ,\nn
\end{equation}
with the infinitesimal gauge parameter $\rho(x)$. To write the $L_\infty$-algebra of classical electrodynamics in a more compact way
we define a master field ${\cal A}\in V_1$ and the corresponding equations of motion $F_{\cal A} \in V_2$ as
\begin{equation}
{\cal A} = \left( \begin{array}{c}
\bar{\psi} \\
\psi\\
A_\mu
\end{array}\right) , \quad F_{\cal A} = \left( \begin{array}{c}
F_{\bar{\psi}} \\
F_\psi\\
(F_A)_\mu
\end{array}\right) . \label{CompositeA_F}
\end{equation}
The corresponding brackets are then
\begin{align} 
&\ell_1 (\rho ) = \left( \begin{array}{c}
0 \\
0 \\
\frac{1}{e}\partial_\mu \rho
\end{array}\right)  ,\quad\quad \ell_1 ({\cal A}) = \left( \begin{array}{c}
i\gamma^\mu\partial_\mu\psi -m\psi \\
-i\gamma^\mu\partial_\mu\bar{\psi} -m\bar{\psi}\\
-\partial_\mu\partial_\nu A^\nu + \partial_\nu\partial^\nu A_\mu
\end{array}\right) ,\nn\\
&\ell _1 (F_{\cal A} ) = \partial_\mu (F_A)^\mu , \quad\quad \ell_2({\cal A}, F_{\cal A}) = -ie( \bar{\psi} F_{\bar{\psi}} - F_\psi \psi) ,\label{ClassEldBrackets}\\
&\ell_2 (\rho, {\cal A} ) = \left( \begin{array}{c}
-i\bar{\psi}\rho \\
i\rho\psi \\
0
\end{array}\right) , \quad  \ell _2 ({\cal A}_1, {\cal A}_2 ) = -\frac{e}{2}\left( \begin{array}{c}
\gamma^\mu A_{1\, \mu} \psi_2 + \gamma^\mu A_{2\, \mu} \psi_1\\
\bar{\psi}_1\gamma^\mu A_{2\, \mu} + \bar{\psi}_2\gamma^\mu A_{1\, \mu}\\
\bar{\psi}_1\gamma^\mu\psi_2 +  \bar{\psi}_2\gamma^\mu\psi_1
\end{array}\right) . \nonumber 
\end{align}
The cyclic pairing can be defined as
\begin{align}\label{CyclPairing} 
\langle \rho, \d_{\cal A}F_{\cal A}\rangle &=\> \int \d^4 x \, \rho \cdot \d_{\cal A}F_{\cal A} \ , \\
\langle {\cal A}, F_{\cal A} \rangle &= \> \int\d^4 x \, \Big( A_\mu F^\mu_A +  F_\psi \psi + \bar{\psi}F_{\bar{\psi}} \Big) \ . \nn
\end{align}

\noindent It is easy to verify that the brackets (\ref{ClassEldBrackets}) and the pairing (\ref{CyclPairing}) reproduce the classical theory:
\begin{align} 
&{\mbox{Gauge transformations:}}\nn\\
&\delta_\rho {\cal A} \>=\>  \left( \begin{array}{c}
\delta_\rho \bar{\psi} \\
\delta_\rho \psi\\
\delta_\rho A_\mu
\end{array}\right) = \ell_1({\rho}) + \ell_2(\rho, {\cal A}) =  \left( \begin{array}{c}
-\ii\,\bar{\psi}\,\rho \\
\ii\,\rho\,\psi \\
\frac1e\,\partial_\mu\rho
\end{array}\right)  \ , \nn\\
&{\mbox{Equations of motion:}}\nn\\
&F_{\cal A} \>=\> \left( \begin{array}{c}
F_{\bar{\psi}} \\
F_\psi\\
(F_A)_\mu
\end{array}\right) = \ell_1({\cal A}) -\frac{1}{2}\, \ell_2({\cal A}, {\cal A}) = \left( \begin{array}{c}
i\gamma^\mu(\partial_\mu\psi -iA_\mu\psi) -m\psi  \\
-i(\partial_\mu\bar{\psi} + i\bar{\psi}A_\mu)\gamma^\mu - m\bar{\psi}  \\
-\partial_\mu\partial_\nu A^\nu + \partial_\nu\partial^\nu A_\mu + e\bar{\psi}\gamma_\mu\psi
\end{array}\right) \ , \nn\\
&{\mbox{Action:}}\nn\\
&S \>=\> \frac{1}{2} \langle {\cal A},\ell_1({\cal A}) \rangle - \frac{1}{3!} \langle {\cal A}, \ell_2({\cal A}, {\cal A})\rangle\nn\\
&= \> \int \d^4x\, \Big( -\frac{1}{4}F^{\mu\nu}F_{\mu\nu} + \bar{\psi}\big( i\gamma^\mu (\partial_\mu\psi -ieA_\mu\psi) -m\psi \big)
\Big)\nn\\
&{\mbox{II Noether identity:}}\nn\\
&\d_\CA F_{\cal A} \>=\> \ell_1(F_{\cal A}) - \ell_2({\cal A}, F_{\cal A}) = \partial_\mu (F_A)^\mu + \ii e\,\bar{\psi}\,F_{\bar{\psi}} - \ii e\,F_\psi\, \psi =0 \ .\nn
\end{align}

\vspace{5mm}
\noindent{\bf $L_\infty$-algebra of braided electrodynamics}
\vspace{3mm}

\noindent Following the steps described in the previous section, we now deform the classical $L_\infty$-algebra of electrodynamics to the braided $L_\infty$-algebra. The corresponding field theory is the braided electrodynamics and it represents a noncommutative deformation of the classical theory. The deformation is introduced by the Moyal-Weyl twist (\ref{MW_Twist}) and the corresponding $\star$-product between functions is given by
\begin{align}
f\star g &=  \>\cdot\circ \Big( e^{\frac{i}{2}\theta^{\mu\nu}\partial_\mu \otimes 
\partial_\nu} f\otimes g \Big) = f\cdot g + \frac{i}{2}\theta^{\mu\nu}\partial_\mu f \cdot \partial_\nu g +\dots ,\label{MW_Star} 
\end{align}
where $\cdot$ represents the usual multiplication.

\noindent The braided brackets are given by:
\begin{align}\label{BraidedEldBrackets}
&\ell^\star_1 (\rho ) = \left( \begin{array}{c}
0 \\
0 \\
\frac{1}{e}\partial_\mu \rho
\end{array}\right)  ,\hspace{2.6cm} \ell^\star_2 (\rho, {\cal A} ) = \left( \begin{array}{c}
-i\sfR_k(\bar{\psi})\star \sfR^k(\rho) \\
i\rho \star \psi \\
i[ \rho, A]_\star =0
\end{array}\right) ,\\
& \ell^\star _1 (F^\star_{\cal A} ) = \partial_\mu (F^\star_A)^\mu , \hspace{2.6cm} \ell^\star_2({\cal A}, F^\star_{\cal A}) = -ie\big( \bar{\psi} \star F_{\bar{\psi}} - \sfR_k (F_\psi)\star \sfR^k (\psi) \big),\nn \\
&\ell^\star_1 ({\cal A}) = \left( \begin{array}{c}
i\gamma^\mu\partial_\mu\psi -m\psi \\
-i\partial_\mu\bar{\psi}\gamma^\mu -m\bar{\psi}\\
-\partial_\mu\partial_\nu A^\nu + \partial_\nu\partial^\nu A_\mu
\end{array}\right),  \quad \ell^\star _2 ({\cal A}_1, {\cal A}_2 ) = -\frac{e}{2}\left( \begin{array}{c}
\gamma^\mu A_{1\, \mu}\star \psi_2 + \sfR_k\gamma^\mu A_{2\, \mu}\star \sfR^k \psi_1\\
\bar{\psi}_1\star \gamma^\mu A_{2\, \mu} + \sfR_k\bar{\psi}_2\star \gamma^\mu \sfR^kA_{1\, \mu}\\
\bar{\psi}_1\gamma^\mu\star \psi_2 +  \sfR_j\bar{\psi}_2\gamma^\mu\star \sfR^j\psi_1
\end{array}\right) , \nonumber 
\end{align}

\noindent The braided electrodynamics is defined with:
\begin{align} 
&{\mbox{Gauge transformations:}}\nn\\
&\delta^\star_\rho {\cal A} \>=\>  \left( \begin{array}{c}
\delta^\star_\rho \bar{\psi} \\
\delta^\star_\rho \psi\\
\delta^\star_\rho A_\mu
\end{array}\right) = \ell^\star_1({\rho}) + \ell^\star_2(\rho, {\cal A}) =  \left( \begin{array}{c}
-\ii\,\sfR_k(\bar{\psi})\star \sfR^k(\rho)  \\
\ii\,\rho\star\psi  \\
\frac1e\,\partial_\mu\rho
\end{array}\right)  \ , \label{BraidedEldGTr}\\
&{\mbox{Equations of motion:}}\nn\\
&F^\star_{\cal A} \>=\> \left( \begin{array}{c}
F^\star_{\bar{\psi}} \\
F^\star_\psi\\
(F_A^\star)_\mu
\end{array}\right) = \ell^\star_1({\cal A}) -\frac{1}{2}\, \ell^\star_2({\cal A}, {\cal A})  = \left( \begin{array}{c}
i\gamma^\mu(\partial_\mu\psi -i e(A_\mu\star \psi + \sfR_k A_\mu\star \sfR^k \psi)) -m\psi  \\
-i(\partial_\mu\bar{\psi} + ie(\bar{\psi}A_\mu + \sfR_k\bar{\psi}\star \sfR^k A_\mu ))\gamma^\mu - m\bar{\psi}  \\
-\partial_\mu\partial_\nu A^\nu + \partial_\nu\partial^\nu A_\mu + \frac{e}{2}\big( \bar{\psi}\star \gamma^\nu\psi + \sfR_k\bar{\psi}\star \gamma^\nu\sfR^k\psi\big)
\end{array}\right) \ , \label{BraidedEldEoM}\\
&{\mbox{Action:}}\nn\\
&S_\star \>=\> \frac{1}{2} \langle {\cal A},\ell^\star_1({\cal A}) \rangle - \frac{1}{3!} \langle {\cal A}, \ell^\star_2({\cal A}, {\cal A})\rangle\nn\\
&= \> \int \d^4x\, \Big( -\frac{1}{4}F^{\mu\nu}\star F_{\mu\nu} + \bar{\psi}\star \big( i\gamma^\mu (\partial_\mu\psi 
-ie A_\mu\star \psi - ie \sfR_k A_\mu\star \sfR^k \psi ) -m\psi \big)
\Big)\label{BraidedEldAction}\\
&{\mbox{II Noether identity:}}\nn\\
&\d_\CA F^\star_{\cal A} \>=\> \ell^\star_1(F_{\cal A}^\star) - \ell^\star_2({\cal A}, F_{\cal A}^\star) = \partial_\mu (F^\star_A)^\mu + \frac{e}{2}\partial_\mu \big( \bar{\psi}\gamma^\mu\star \psi + \sfR_k(\bar{\psi})\gamma^\mu \star \sfR^k(\psi)\big) =0 \, .\label{BraidedEldNoether}
\end{align}

\noindent The second Noether identity, combined with the equations of motion can be used to derive the conserved charge of the braided $U(1)$ gauge theory. Setting $F_A^\star =0$ in (\ref{BraidedEldNoether}) leads to
\begin{equation}
\frac{e}{2}\partial_\mu \big( \bar{\psi}\gamma^\mu\star \psi + \sfR_k(\bar{\psi})\gamma^\mu \star \sfR^k(\psi)\big) =0 ,\nn 
\end{equation}
that is the matter current is conserved. The corresponding conserved charge is then given by
\begin{equation}
Q^\star =  \frac{e}{2}\int_B {\rm d} ^3\vec{x}\, \big( \psi^\dagger\star \psi + \sfR_k(\psi^\dagger)\star \sfR^k(\psi)\big). \label{BraidedEldCharge}
\end{equation}
Although the Moyal--Weyl $\star$-product is compatible with the cyclic inner product \eqref{CyclPairingStar}
\begin{equation}
\int\, \d^4 x\ \sfR_k(f)\star \sfR^k(g) = \int\, \d^4 x\ f\star g \ ,\nn
\end{equation}
this cannot be applied to the integration in (\ref{BraidedEldCharge}) which is only taken over the three-dimensional spatial volume $B$. Hence the second term in (\ref{BraidedEldCharge}) has a non-trivial contribution to the conserved charge, and $Q^\star$ generally differs from the electromagnetic charge not only in the classical theory but also in the $\star$-deformed electrodynamics. Only when the time component of the twist vanishes, that is $\theta^{0i}=0$, do we recover the usual electromagnetic charge.

\vspace{5mm}
\noindent{\bf A first look at quantization}
\vspace{3mm}

\noindent The usual starting point for quantization of a field theory is the classical action. In the case of braided electrodynamics, we can start from (\ref{BraidedEldAction}) and split it into three pieces
\begin{align}
S_\star & = \int \d^4x\, \Big\{ -\frac{1}{4}F^{\mu\nu}\star F_{\mu\nu} + \bar{\psi} \star i\gamma^\mu \partial_\mu\psi 
+ \frac{e}{2}\Big(\bar{\psi}\star A_\mu \gamma^\mu\star\psi + \bar{\psi}\star \sfR_k(A_\mu) \gamma^\mu\star\sfR^k(\psi)
\Big\} \nn\\
&= \int \d^4x\, \Big\{ \mathcal{L}_A + \mathcal{L}_{\psi} + \mathcal{L}_{\mbox{\tiny int}}\Big\} . \label{MWLagranzijani}
\end{align}

\noindent Since the action (\ref{MWLagranzijani}) is invariant under the braided $U(1)$ gauge symmetry, we have to perform gauge fixing before the quantization. The ghost field that appears in this process completely decouples from the gauge field $A$ due to the abelian nature of the braided $U(1)$ symmetry. The full action is then given by
\begin{equation}
S_\star =  \int \d^4x\, \Big\{ \mathcal{L}_A + \mathcal{L}_{\psi} + \mathcal{L}_{\mbox{\tiny int}} + \mathcal{L}_{\mbox{\tiny ghost}}\Big\} ,\label{MWQFTLagranzijani}
\end{equation}
with $\mathcal{L}_{\mbox{\tiny ghost}} = -\bar{c}\star \partial_\mu\partial^\mu c$. The corresponding vertex is given in Figure 1.

\begin{figure}[h]
\centering
\includegraphics[scale=0.15]{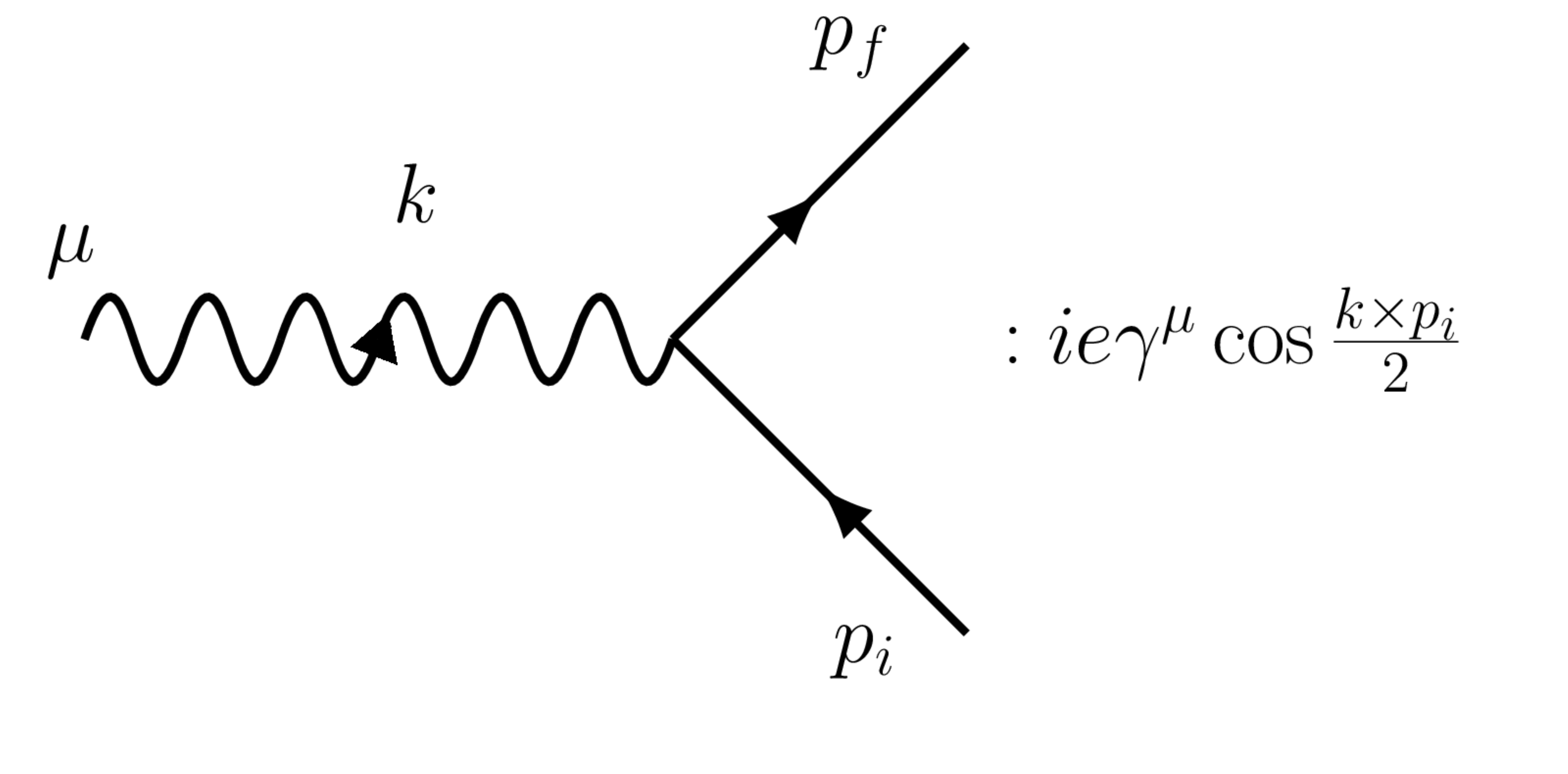}
\caption{Photon-fermion vertex in braided electrodynamics.}\label{Verteks}
\end{figure}

\noindent The notation $a\times b=a_\mu b_\nu\theta^{\mu\nu}$ is used. As expected, the propagators do not change compared with the undeformed case, while the vertex has a nontrivial contribution due to the noncommutative deformation. In the $\star$-deformed electrodynamics the noncommutative correction to the photon-fermion vertex consists only of the phase factor; in the braided electrodynamics, due to the presence of the $\RR$ matrix in the action (\ref{MWLagranzijani}), the phases combine to the cosine factor. Note that unlike in the $\star$-deformed electrodynamics, here there are no three and four photon vertices and no photon-ghost vertex.

\begin{figure}[h]
\centering
\includegraphics[scale=0.7]{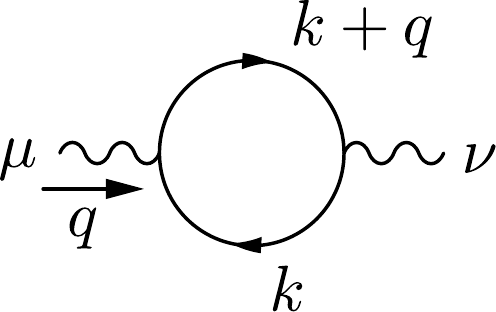}
\caption{Photon self-energy.}\label{VacPol}
\end{figure}

\noindent As an illustration, let us calculate the photon self-energy at one loop order. The only contribution comes from the fermion bubble in Figure \ref{VacPol}.

\noindent The corresponding amplitude is
\begin{equation}
i\Pi_2^{\mu\nu}(q)=(-1)(-ie)^2\int \frac{\d ^4k}{(2\pi)^4}\Tr(\gamma^\mu\frac{i}{\slashed{k}-m}\gamma^\nu\frac{i}{\slashed{k}+\slashed{q}-m})\cos^2{(\frac{k\times q}{2})}. \nn
\end{equation}
After a straightforward calculation\footnote{We use dimensional regularization and set $D=4-\epsilon$.} following the appendix B, we obtain
\begin{eqnarray}\label{eq:PiNCIzracunat}
i\Pi_2^{\mu\nu}(q) &=& -\frac{ie^2}{(2\pi^2)}\int_0^1 \d x\, \left( \frac{4\pi\mu^2}{\Delta}\right)^\frac{\epsilon}{2}
(\eta^{\mu\nu}q^2 - q^\mu q^\nu)x(1-x)\Gamma(\frac{\epsilon}{2}) \nn\\
&&-\frac{ie^2}{(2\pi^2)}\int_0^1 \d x\, \left( \frac{4\pi\mu^2}{\Delta}\right)^\frac{\epsilon}{2} \label{PiCeoIzracunat}\\
&& \Bigg\{ (\eta^{\mu\nu}q^2 - q^\mu q^\nu)x(1-x)\Big( \big(\frac{|\bar{\theta}|\sqrt{\Delta}}{2}\big)^{\frac{\epsilon}{2}}K_\frac{\epsilon}{2}(|\bar{\theta}|\sqrt{\Delta}) -\frac{1}{2}\Gamma(\frac{\epsilon}{2}) \Big)\nn\\
&&\hspace{5mm} -\Delta \frac{\bar{\theta}^\mu\bar{\theta}^\nu}{\bar{\theta}^2}\Big(
(1-\frac{D}{2})\big(\frac{|\bar{\theta}|\sqrt{\Delta}}{2}\big)^{1-\frac{D}{2}}K_{1-\frac{D}{2}}(|\bar{\theta}|\sqrt{\Delta})
- \big(\frac{|\bar{\theta}|\sqrt{\Delta}}{2}\big)^{\frac{\epsilon}{2}}K_\frac{\epsilon}{2}(|\bar{\theta}|\sqrt{\Delta}) \Big)
\Bigg\} .\nn
\end{eqnarray}
Here we introduced $\bar{\theta}^\mu=\theta^{\mu\nu}q_\nu$, $|\bar{\theta}|=\sqrt{-\bar{\theta}^\mu\bar{\theta}_\mu}$ and $\Delta = m^2 - x(1-x)q^2$. The parameter $\mu$ has mass dimension $1$ and is introduced for dimensional reasons.

\noindent The first line of (\ref{PiCeoIzracunat}) is the usual commutative result. Expanding $\Gamma(\frac{\epsilon}{2})$ as $\Gamma(\frac{\epsilon}{2})\approx \frac{2}{\epsilon} -\gamma + {\cal O}(\epsilon)$ and 
\begin{equation}
\left( \frac{4\pi\mu^2}{\Delta}\right)^\frac{\epsilon}{2}\approx 1 + \frac{\epsilon}{2}\ln \frac{4\pi\mu^2}{\Delta} \nn
\end{equation}
we find the usual divergence in the commutative contribution
\begin{equation}
i\Pi_2^{\mu\nu}(q)\Big|_{\theta =0} \approx  -\frac{ie^2}{(2\pi^2)}(\eta^{\mu\nu}q^2 - q^\mu q^\nu)\Big( \frac{1}{3\epsilon} - \frac{\gamma}{6} + \int_0^1 \d x\, x(1-x) \ln \frac{4\pi\mu^2}{m^2-q^2x(1-x)} + {\cal O}(\epsilon)\Big).\label{PiCommDivergencija}
\end{equation}
The noncommutative part of the result (\ref{eq:PiNCIzracunat}) is convergent \cite{U1UsefulFormula}. However, in the limit $|\bar{\theta}|\rightarrow 0$ ($\theta^{\mu\nu}=0$ or $q\rightarrow 0$) it becomes UV divergent \cite{U1UsefulFormula}. We conclude that in this way we recover the UV/IR mixing, despite the facts that there are no nonplanar diagrams. This unexpected result simply reflects our current lack of understanding of the proper quantization of field theories with braided symmetry, which is currently under development \cite{UsNew}. Some examples \cite{Oeckl, SzaboAlex} suggest that a proper quantization of braided (gauge) field theories results in the absence of non-planar diagrams and at the same time the absence of UV/IR mixing. 

\noindent Analyzing further the noncommutative contribution (\ref{eq:PiNCIzracunat}) we find that it does not spoil the transversality of the photon. The second line of (\ref{eq:PiNCIzracunat}) is obviously transversal. To see that the third line is also transversal we multiply it with $q^\mu$ and find
\begin{equation}
q_\mu \frac{\bar{\theta}^\mu\bar{\theta}^\nu}{\bar{\theta}^2}\Big(\dots\Big) =q_\mu \frac{\theta^{\mu\lambda}q_\lambda\theta^{\nu\sigma}q_\sigma}{\bar{\theta}^2}\Big(\dots\Big) =0 \nn
\end{equation}
since $\theta^{\mu\lambda}q_\lambda q_\mu=0$ due to the antisymmetry of $\theta^{\mu\lambda}$.

\noindent As a comparison, we mention that in the $\star$-deformed electrodynamics there are three more diagrams contributing to the $\Pi_2^{\mu\nu}(q)$: the photon bubble, the photon tadpole and the ghost bubble (ghosts do not decouple from the photon due to the nonabelian nature of the $\star$-deformed electrodynamics). These three diagrams give non-trivial corrections to the commutative result, while the fermion bubble gives no noncommutative corrections. The UV/IR mixing is present \cite{U1ChargeQuantization, U1UsefulFormula}.

\section{Outlook}

\noindent In this paper we applied the recently developed formalism of braided $L_\infty$-algebras to construct a noncommutative deformation of the classical electrodynamics. The obtained theory, the braided electrodynamics is invariant under the braided $U(1)$ gauge symmetry, which is still abelian. As a consequence, there are no three and four photon interaction vertices in the action $S_\star$ (\ref{BraidedEldAction}). Therefore, the only interaction vertex is the fermion-photon vertex. This is different compared to the $\star$-gauge deformation of the classical electrodynamics which is nonabelian and as a consequence the three and four photon interaction vertices appear.\smallskip

\noindent We calculated the Feynman integrals which appear in the one-loop contribution to the vacuum polarization. Although there are no non-planar diagrams contributions, we find the UV/IR mixing. This unexpected result is consistent with claims about equivalence of various combinations of (non)commutative field theories and (un)braided statistics made in \cite{Oeckl}.\smallskip

\noindent The correct quantization of theories with braided symmetries should be implemented with braided homotopy algebraic techniques, and will be addressed in \cite{UsNew}. The results of \cite{SzaboAlex, Oeckl} suggest that there is no UV/IR mixing, at least in the scalar field theories which do not possess gauge symmetries. The results reported in this preliminary investigation suggest that there are still many new physical features to be uncovered in braided quantum field theory.

\appendix

\renewcommand{\theequation}{\Alph{section}.\arabic{equation}}
\setcounter{equation}{0}

\section{Drinfel'd twist deformation}
\label{app:Drinfeld}

\noindent In this appendix we briefly introduce the basics of Drinfel'd deformation and the notation we use on the example of the Moyal-Weyl deformation. More details can be found in \cite{MajidBook, SLN}.

\noindent In the Drinfel'd twist formalism, a deformation is introduced by twist ${\cal F} \in U\frv \otimes U\frv$ 
\begin{equation}
{\cal F} =  e^{-\frac{i}{2}\theta^{\mu\nu}\partial_\mu \otimes 
\partial_\nu} ,\label{MW_Twist} 
\end{equation}
where $\theta^{\mu\nu}$ is a constant antisymmetric matrix and its entries are 
considered to
be small deformation parameters\footnote{To be more precise, the twist (\ref{MW_Twist}) should be written as
\begin{equation}
e^{-\frac{i}{2}\kbar\theta^{\mu\nu}\partial_\mu \otimes 
\partial_\nu}
 \end{equation}      
with the small deformation parameter $\kbar$ and arbitrary constant 
antisymmetric matrix elements $\theta^{\mu\nu}$. In the usual notation $\kbar$ is 
absorbed in the matrix elements $\theta^{\mu\nu}$ and these are called 
deformation parameters.
}. Note that $\partial_\mu$ belong to the Lie
algebra of vector fields $\frv:=\Gamma(TM)$ on a manifold $M$. The corresponding enveloping algebra is $U\frv$. We use the following notation: ${\cal F} = {\rm{f}}^k\otimes {\rm{f}}_k$, ${\cal F}^{-1} =  \bar{\rm{f}}^k\otimes 
\bar{\rm{f}}_k$. 

\noindent The invertible $\RR$-matrix $\RR\in
U\frv\otimes U\frv$ encodes the braiding (deformation, noncommutativity) and it is induced by the twist as
\begin{align}
\RR=\mathcal{F}_{21}\, {\cal F}^{-1}=:\sfR^k\otimes\sfR_k \ ,
\end{align}
where $\mathcal{F}_{21}=\tau(\mathcal{F})=\mathrm{f}_k\otimes\mathrm{f}^k$ is the twist with its legs swapped. It is easy to see that the $\RR$-matrix is triangular, that is
\begin{equation}
  \RR_{21} = \RR^{-1} = \sfR_k\otimes\sfR^k \ .\label{Inv_R}
\end{equation}

\noindent The twist (\ref{MW_Twist}) deforms the algebra of functions $C^\infty(M)$ to $C^\infty(M)_\star$
\begin{align}
f\star g &=\> \bar{\rm{f}}^k(f)\cdot \bar{\rm{f}}_k(g) \label{fstarg} = \sfR_k\, g\star \sfR^k\, f\\ 
&=\> f\cdot g + \frac{i}{2}\theta^{\mu\nu}\partial_\mu f \cdot \partial_\nu g + \dots \ .  \nn 
\end{align}

\noindent The exterior algebra of differential forms $\Omega^\sharp(M)$ is deformed in a similar way to $\Omega^\sharp(M)_\star$ with
\begin{align}
& \omega_1\wedge_\star\omega_2 = \bar{\rm{f}}^k(\omega_1)\wedge \bar{\rm{f}}_k(\omega_2) = (-1)^{d_1d_2}\sfR_k\,\omega_2\wedge_\star \sfR^k\, \omega_1,\label{StarWedge} \\
& \d(\omega_1\wedge_\star\omega_2) = \d\omega_1\wedge_\star\omega_2 + (-1)^{d_1}\,
\omega_1\wedge_\star\d \omega_2  \ .\nn
\end{align}
The exterior derivative $\d$ is undeformed and the degrees of forms $\omega_1, \omega_2$ we label with $d_1, d_2$ respectively.

\section{Modified Bessel functions of the second kind}

\noindent The differential equation
\begin{equation}
\frac{\d ^2 u}{\d z^2} + \frac{1}{z}\frac{\d u}{\d z} - (1+ \frac{\nu^2}{z^2})u =0 
\end{equation}
has a solution
\begin{equation}
u = C_1 I_\nu(z) + C_2 K_\nu(z) 
\end{equation}
where $I_\nu(z)$ and $K_\nu(z)$ are modified Bessel functions of the first and the second kind \cite{GrRTablice}.

If $n+1$ is a natural number, then $K_n(z)$ has the following series expansion
\begin{eqnarray}
K_n(z) &=& \frac{1}{2}\sum_{k=0}^{n-1} (-1)^k\frac{(n-k-1)!}{k!\left(\frac{z}{2}\right)^{n-2k}}\nn\\
&& + (-1)^{n+1}\sum_{k=0}^{\infty}\frac{\left(\frac{z}{2}\right)^{n+2k}}{k!(n+k)!}\Big( \log \frac{z}{2} - \frac{1}{2}\psi(k+1)- \frac{1}{2}\psi(n+k+1)\Big) ,\label{C2}
\end{eqnarray}
where $\psi(k)$ is the Euler psi function
\begin{equation}
\psi(k) = -\frac{1}{k}-\gamma-\sum_{l=1}^\infty \big( \frac{1}{x+l} - \frac{1}{l}\big)\nn
\end{equation}
and $\gamma$ is the Euler constant.
We also use the limit
\begin{equation}
\lim_{z\to 0} z^\nu K_\nu (z) = 2^{\nu-1}\Gamma(\nu) .\label{C3}
\end{equation}

\noindent
It was shown in \cite{U1UsefulFormula} that

{\small
\begin{eqnarray}
&&\int \d^D l_{\mbox {\tiny E}}\frac{e^{il_{\mbox {\tiny E}}\bar{\theta}}}{(l_{\mbox {\tiny E}}^2 + \Delta)^\alpha} = \frac{(-1)^\alpha 2\pi^{\frac{D}{2}}}{\Delta^{\alpha-\frac{D}{2}}\Gamma(\alpha)}(\frac{|\bar{\theta}|\sqrt{\Delta}}{2})^{\alpha-\frac{D}{2}}K_{\alpha-\frac{D}{2}}(|\bar{\theta}|\sqrt{\Delta}), \label{UF1}\\
&&\int \d^D l_{\mbox {\tiny E}}\frac{l_{\mbox {\tiny E}}^\mu l_{\mbox {\tiny E}}^\nu\quad e^{il_{\mbox {\tiny E}}\bar{\theta}}}{(l_{\mbox {\tiny E}}^2+\Delta)^\alpha} = \frac{(-1)^\alpha \pi^{\frac{D}{2}}}{\Delta^{\alpha-1-\frac{D}{2}}\Gamma(\alpha)}\Bigg[(\frac{|\bar{\theta}|\sqrt{\Delta}}{2})^{\alpha-1-\frac{D}{2}}K_{\alpha-1-\frac{D}{2}}(|\bar{\theta}|\sqrt{\Delta})\eta^{\mu\nu} \label{UF2}\\
&& + \frac{\bar{\theta}^\mu\bar{\theta}^\nu}{\bar{\theta}^2}\Big((2\alpha-2-D)(\frac{|\bar{\theta}|\sqrt{\Delta}}{2})^{\alpha-1-\frac{D}{2}}K_{\alpha-1-\frac{D}{2}}(|\bar{\theta}|\sqrt{\Delta})-2(\frac{|\bar{\theta}|\sqrt{\Delta}}{2})^{\alpha-\frac{D}{2}}K_{\alpha-\frac{D}{2}}(|\bar{\theta}|\sqrt{\Delta})\Big)\Bigg]. \nn
\end{eqnarray}
}
The integration is done in the Euclidean momentum space and $\bar{\theta}^\mu=\theta^{\mu\nu}q_\nu$, $|\bar{\theta}|=\sqrt{-\bar{\theta}^\mu\bar{\theta}_\mu}$.

\paragraph{Acknowledgments.}
We thank the organisors of the Corfu Summer Institute
2021 for the stimulating meeting and the opportunity to present the
preliminary results of our work.
The work of {\sc M.D.C., N.K.} and  {\sc V.R.} is supported by Project
451-03-9/2021-14/200162 of the Serbian Ministry of Education, Science and
Technological Development. The work of {\sc R.J.S.} was supported by
the Consolidated Grant ST/P000363/1 
from the UK Science and Technology Facilities Council.

\end{document}